\documentstyle[11pt,inaoe_proc,psfig2,twoside]{article} 
\def\apj{ApJ}

\def\BP{Ballesteros-Paredes}

\def\eikx{e^{i \k \cdot \x}}
\def\F{{\cal F}}
\def\k{{\bf k}}

\def\p{{\bf p}}
\def\q{{\bf q}}
\def\r{{\bf r}}

\def\u{{\bf u}}
\def\uk{{\u_\k}}
\def\va{v_{\rm A}}
\def\VS{V\'azquez-Semadeni}
\def\x{{\bf x}}

\begin{document}

\title{Turbulence in Molecular Clouds}  

\author{Enrique V\'azquez-Semadeni} 
\affil{ Instituto de Astronom\'\i a, UNAM\\
Apartado Postal 70-264, M\'exico D.F. MEXICO  }  

\begin{abstract}
In this course we review the  theory of incompressible homogeneous
turbulence at an elementary level, and discuss the similarities and
differences expected in the compressible case, relevant to the
interstellar medium and molecular clouds. We stress that a
general definition of turbulence applicable to the compressible case
should not rely on the Kolmogorov $k^{-5/3}$ spectrum nor on an energy
cascade from large to small scales. Instead, we discuss the
various possibilities for 
the energy spectrum of compressible turbulence, which numerical
simulations suggest should be $\sim k^{-2}$, and the nature of the cascades,
if at all present.
We then discuss issues concerning molecular clouds which are
likely to be directly related to turbulence, such as cloud formation,
cloud structure, and cloud support against gravity.
\end{abstract}


To appear in ``Millimetric and Sub-Millimetric Astronomy. INAOE 1996 Summer
School''.

\section{ Introduction }
\markboth{Enrique V\'azquez-Semadeni}{Turbulence in Molecular Clouds}

In recent years, a wide variety of scientific disciplines have come to
the realization that nonlinear phenomena
are the norm rather than the exception. Most physical,
astronomical, biological, social and economic systems are strongly
nonlinear, and in fact the time is ripe for a change in our
categorization of such systems. While our current classifications of
dynamical behavior are based on a ``linear vs.\ nonlinear'' scheme,
intrinsically attributing preponderance to linear systems, a ``complex vs.\
non-complex'' classification would probably be more appropriate,
reflecting the fact that most systems in Nature and society are
complex. 

Turbulent flows are a prime example of complex systems, and it is
well known that a complete theory even of incompressible turbulence
does not exist. By this it is meant that some specific statistical
properties of turbulent flows, such as the energy spectrum and the
higher order correlation functions, cannot be derived directly from the
equations of motion without the introduction of simplifying
assumptions. Nevertheless, phenomenological theories exist for  both
three-dimensional (Kolmogorov 1941) and two-dimensional (Kraichnan
1967) turbulence in
incompressible flows, which have been extremely influential. In
fact, the Kolmogorov theory has almost become synonymous with
turbulence. However, it will be argued in this course that such an
identification is likely not to be adequate for compressible
turbulence, and thus a more
appropriate definition of turbulence is that {\it a)} it contains an
extremely large number of excited degrees of freedom (or modes); {\it b)}
the modes are able to nonlinearly exchange excitation; {\it c)} the
system is unpredictable in the sense of exhibiting {\it sensitivity to
initial conditions}, and {\it d)} the system is {\it mixing} 
(see, e.g., Scalo 1987; Lesieur 1990; Frisch 1995). 
Property c) is a distinctive feature of {\it chaotic} systems. 

The interstellar medium (ISM) in particular is an extremely complex
system, including gaseous, dust, cosmic ray and magnetic field
components in a strongly turbulent state. 
As discussed in the Virial Theorem chapter of this book,
the gaseous and magnetic components of the ISM are probably well
described by the magnetohydrodynamic (MHD) equations (see, e..g.,
Cowling 1976; Spitzer 1978; Shore 1992; Shu 1992). In this course, we
will discuss some basic theoretical aspects of compressible
turbulence, remarking its differences with the incompressible case (\S\
\ref{theory}), and then review some observational and theoretical work
regarding several aspects of turbulence and molecular clouds, namely
cloud formation, cloud structure and cloud support and star formation
(\S\ \ref{applications}), emphasizing on recent
developments. Extensive reviews of earlier work can be found in
Dickman (1985) and Scalo (1987). Finally, \S 4 presents a summary and
conclusions.

\section{Basic Theory}\label{theory}

\subsection{Incompressible hydrodynamic
  turbulence}\label{incompressible}

In this section we discuss the basic notions necessary for the subsequent
discussions. Comprehensive presentations of homogeneous turbulence can be
found in Batchelor (1953), Tennekes \& Lumley (1972), Rose \& Sulem
(1978) and Lesieur (1990).

\subsubsection{The Navier-Stokes equation.}

The full MHD equations have already been presented in the chapter on
the Virial Theorem. Here we just discuss the incompressible
momentum conservation equation without the Lorentz, Coriolis, or other external
forces. When including the constitutive
relation for stress in a newtonian fluid (see, e.g., Currie 1974) the
momentum equation is
referred to as the Navier-Stokes (NS) equation. It reads:
\begin{equation}\label{NSeq}
{\partial\u\over\partial t} + \u\cdot\nabla\u =
-{\nabla P \over \rho} +\nu \nabla^2 \u,
\end{equation}
where \u\ is the velocity field, $\rho$ is the density, $P$ is the
pressure, and $\nu$ is the {\it kinematic viscosity}. Additionally,
the incompressibility condition $d \rho/ dt=0$ applied to the
continuity (or mass conservation) equation yields
\begin{equation}\label{incompressibility}
\nabla \cdot \u=0.
\end{equation}

As mentioned in the Virial Theorem chapter, the second term in the LHS
of eq.\ (\ref{NSeq}) is called the {\it advective} or nonlinear term,
and is the one responsible for turbulent transport of momentum (the
$j$-th component of the velocity transporting the $i$-th component of
the momentum). On the other hand, viscosity tends to unify the motion
of fluid parcels. Thus, a measure of the ``amount of turbulence''
present in the flow is given by the ratio of the advective to the
viscous term, known as the {\it Reynolds} number $Re$. Dimensionally,
\begin{equation}\label{reynolds}
Re \equiv {|\u \cdot \nabla \u| \over |\nu \nabla^2 \u|} \sim {U L \over \nu},
\end{equation}
where $U$ is a characteristic value of the velocity, and $L$ is the
characteristic length scale of the flow.

\subsubsection{Fourier decomposition.}\label{fourier}

Much of turbulence theory is done in Fourier space, as it permits a
direct treatment of length {\it scales} -- the wavelengths associated
to the Fourier modes. We take a Fourier decomposition of the velocity
field of the form $\u(\x,t)=\int u_\k(t)\eikx d^3k$, where \k\ is the
vector Fourier mode, with associated wavenumber $k=2 \pi /\lambda$,
$\lambda$ is the mode wavelength, and $\u_\k$ is the
corresponding mode amplitude. 
Under the assumption of periodic boundary
conditions in a domain of size $L$, the wavevectors \k\ have discrete
components, multiples of $2 \pi/L$. 
Choosing a box length $L=2 \pi$ thus
gives wavevectors with integer components. In Fourier space, the NS
equation becomes
\begin{equation}\label{NSFSeq}
{d\uk \over dt} + i \sum_{{\p}+{\q}=\k} \Bigl[(\k \cdot \u_\p)\u_\q -
{\k \over k^2} (\k \cdot \u_\p)(\k \cdot \u_\q)\Bigr] = - \nu k^2\u_\k
\end{equation}
with the incompressibility condition
\begin{equation}\label{incompFS}
\k \cdot \u_\k =0.
\end{equation}

Several points are worth noting. The first term in the brackets
is the advective term. The second term in the brackets is the pressure
gradient term. Interestingly, in the incompressible case, the pressure
gradient term can be solved for in terms of the velocity field in
Fourier space. This is because the incompressible equations behave as
if the speed of sound were infinite, short-circuiting the longer
interaction path
for the compressible case, in which compressive motions cause density
fluctuations via the $\nabla \cdot \u$ term in the continuity
equation, which in turn cause pressure fluctuations (mediated by the
speed of sound), which finally feed back in the momentum equation via
the $\nabla P$ term. 

Secondly, both the advective and the pressure terms are seen to be nonlinear,
i.e., to involve products of pairs of velocity modes. Moreover, it is
seen that all modes \p\ and \q\ such that $\p + \q = \k$
enter the RHS of the equation for mode \k,
so that energy (or, more generally, {\it excitation})
is exchanged among all Fourier modes. This process is
known as {\it nonlinear transfer}. The term excitation is used to
include transfer among modes of different kinds (density, velocity,
internal energy, etc.), which will be appropriate for the compressible
case.

Third, note that eq.\ (\ref{NSFSeq}) actually represents an enormously
large number of coupled {\it ordinary} differential equations (ODEs) for the
Fourier amplitudes $\u_\k$. For example, if we were to perform a
numerical simulation with 1000 grid points per dimension in three
dimensions, this would amount to roughly $10^9$ Fourier modes, or {\it
degrees of freedom}. Although very large for current computer
capabilities, this is still a very small resolution for real
problems. It was pointed out in the Virial Theorem chapter that there
is at least a factor of $10^5$ between the molecular mean free path
and the characteristic scales of the system in most ISM structures,
implying at least $10^{15}$ degrees of freedom (or equations, in the
Fourier description). Thus, the Fourier description trades the
complexity of a single partial differential equation for a very large
number of ODEs.

Fourth, although not readily seen from the equations, turbulent flows
generally exhibit {\it sensitivity to initial conditions}, i.e.,
infinitesimal changes in the initial conditions can lead to markedly
different states after finite times (see, e.g., Ott 1993).

Finally, note that the incompressibility condition clearly
shows that {\it no longitudinal modes exist in incompressible flow}. 

\subsubsection{Energy spectrum and correlation function.} \label{spectrum}

The Fourier description leads naturally to the definition of an
important quantity in turbulence theory, namely the {\it energy
spectrum} $E(k)$. This is defined as the kinetic energy (per unit mass)
contained in modes with wavenumbers in the interval
$(k-1/2,k+1/2]$. This amounts to summing up the energies
$|\u_\k|^2$ of all modes in a spherical shell of thickness $\Delta k
=1$ and of radius $k$. Mathematically,
\begin{eqnarray}\label{spectrumeq}
E &\equiv & {1 \over 2} \int |\u(\x)|^2 d^3x \nonumber \\
  & & = {1 \over 2} \int |\u_\k|^2 d^3k \nonumber \\
  & & = {1 \over 2} \int_0^\infty E(k) dk,
\end {eqnarray}
where the second equality follows from the well-known result for
complex variables, Parseval's relation. Thus
\begin{equation}
E(k) = {1 \over 2} \int \int |\u_\k|^2 k^2 \sin \theta_k d \theta_k d \phi_k.
\end{equation}
The exponent of $k$ in the spectrum is commonly referred to as the
spectral {\it index} or {\it slope}, the latter referring to the slope
of the spectrum in a log-log plot.

It can be readily shown that the energy spectrum is actually the
Fourier transform of the so-called {\it auto-correlation
  function}. This is defined as $C(\r)\equiv \langle \u(\x) \cdot \u(\x +
\r) \rangle/\sigma^2$, where $\sigma$ is the velocity variance.
This function measures the probability that two points separated
by a distance \r\ have the same velocity. A characteristic length is
the {\it correlation length}, which is the $e$-folding distance for
the correlation function. Note that the correlation function can also
be defined for the density field, as in Cosmology, or for products of
different variables, in which case it is called a ``cross
correlation''. A related quantity, the {\it structure function}, is
defined as $S(\r)\equiv \langle |\u(\x) - \u(\x + \r)|^2 \rangle/\sigma^2$.

\subsubsection{Kolmogorov-Obukhov theory.}

In 1941, Kolmogorov and Obukhov independently derived the expected
scalings of all powers of the structure functions as functions of 
separation $r$ for {\it homogeneous} turbulence (i.e.,
whose statistical properties are independent of position). From there,
the form of the energy spectrum readiliy follows. This result
is a milestone in turbulence theory, and is often referred to as the
K41 theory. It is based on the following assumptions:

\smallskip
\noindent 
i) The Reynolds number is large enough that a very large range of scales are
active in the flow.

\smallskip
\noindent 
ii) Energy is injected primarily at large scales and dissipated at
small scales (as can be seen from the $k^2$ dependence of the viscous
term in eq.\ [\ref{NSFSeq}]).

\smallskip
\noindent 
iii) Energy transfer is {\it local} in Fourier space, i.e., it occurs
mainly among similar wavenumbers. Together with the two previous
assumptions, this implies that there is an intermediate range of
wavenumbers in which energy can only {\it cascade} from large to small
scales (small to large $k$). This is called the {\it inertial range}
of wavenumbers.

\smallskip
\noindent 
iv) The system is in statistical equilibrium, so that the rates of
energy injection at large scales, of transfer at intermediate scales
and dissipation at small scales are equal. In consequence, the
transfer rate in the inertial range is independent of wavenumber $k$.

With these assumptions, the form of the spectrum in the inertial range
can be estimated from dimensional analysis as follows. The energy
transfer rate for {\it eddies} (vortices) of scale size $l$ 
can be written as $\epsilon \sim V_l^2/\tau_l$, where
$v_l$ is the characteristic velocity difference accross the eddy,  
and $\tau_l = l/v_l$ is the characteristic or {\it eddy turnover} time at scale
$l$. Thus, the velocity differences scale with separation $l$ as
\begin{equation}\label{veldif}
v_l = (\epsilon l)^{1/3}. 
\end{equation}
Now, the characteristic velocity difference at scale $l$ is related to
the energy spectrum by 
\begin{equation}
v_l^2 = 2 \int_{2 \pi/l}^\infty E(k) dk.
\end{equation}
Assuming that the
spectrum has a power-law dependence on wavenumber $k^{-n}$, and substituting
the scaling relation (\ref{veldif}), we obtain $l^{2/3} \sim \int_{2
  \pi/l}^\infty k^{-n} dk$, implying $n=5/3$. This is the
celebrated Kolmogorov ``$-5/3$ law'', and has been verified
experimentally (e.g., Grant et al.\ 1962). A more detailed discussion of
this phenomenological derivation can be found in Rose \& Sulem (1978).

In summary, the K41 theory brings in two fundamental concepts, which
sometimes are even considered as the trademarks of turbulent flows: a
$k^{-5/3}$ energy spectrum and the concept of an energy cascade from
large to small scales. We will see in \S \ref{compressible}, however, that
compressible turbulence is not necessarily characterized by these
properties. Furthermore, it should also be stressed that even for
incompressible flows the K41 theory does not always hold. For example,
for sheared flows, Malerud et al. (1995) have found that the structure
function exhibits power laws different from that expected in the K41
theory. Thus, K41 theory is only expected to apply when the assumptions
of incompressibility, 
homogeneity, forcing at large scales, large Reynolds number (large
range of excited scales) and statistical equilibrium are satisfied.

Finally, it should be remarked that the energy spectrum loses a
considerable amount of information due to the averaging over angles in
$k$-space. Armi \& Flament (1985) have shown that the
truly relevant information concerning the structure of turbulent flows
resides not in the energy spectrum, but in the complex phases of the
Fourier modes, which are lost upon angle-averaging. They
accomplished this by considering an ocean image, taking its Fourier
transform, performing alterations on the Fourier modes, and then
inverse-transforming the image. When the alterations amounted to
changing the spectral slope of the field, only minor diferences in
contrast were obtained. However, when the alterations amounted to
scrambling the phases of the modes in a random manner, all the
structural information of the image was lost. Thus,
the energy spectrum is by no means completely determinant of the type of flow
present in the fluid.

\subsubsection{Two-dimensional turbulence.}

Two-dimensional (2D) turbulence has received comparable attention to
the three-dimensional (3D) case among the turbulence community 
for a number of reasons. First, the large-scale motions of the Earth's
oceans and athmosphere are very nearly 2D. Second, resolutions large enough
to obtain fully developed turbulence in numerical simulations had
only been achieved in 2D until very recently. Finally, 2D turbulence
exhibits distinctive physical properties which constitute a
challenging theoretical problem on its own right. Exhaustive
discussions on 2D turbulence can be found, for example, 
in Kraichnan \& Montgomery (1980) and Lesieur (1990).

For 2D turbulence, the shape of the energy spectrum was studied by
Kraichnan (1967), on the basis of a dimensional argument similar
to Kolmogorov's for the cascade of {\it enstrophy}, the mean
squared vorticity of the flow. The vorticity is defined as $\omega
\equiv \nabla \times \u$ and the enstrophy as $\Omega \equiv (1/2)\int
\omega^2 dV$. Assuming that energy is injected into the system at
wavenumbers $k_f$ and that wavenumbers both larger and smaller than
$k_f$ are present, Kraichnan predicted that in fact two different spectral
slopes should be found in 2D: a $-3$ slope at wavenumbers larger than
$k_f$ (scales smaller than $2 \pi/k_f$), and a $-5/3$ slope at
wavenumbers smaller than $k_f$ (scales larger than $2
\pi/k_f$). Furthermore, he predicted that a {\it direct} (from larger
to smaller scales) cascade of
{\it enstrophy} should occur in the $k^{-3}$ range, while an {\it
  inverse} (from smaller to larger scales) cascade of {\it energy} should be
present in the $k^{-5/3}$ range. These predictions have been verified
numerically (Herring et al 1974; Frisch \& Sulem 1986), and constitute
the standard reference against which to compare numerical results in the
2D compressible case, equivalent to the Kolmogorov spectrum in the 3D
case.

It should nevertheless be pointed out that the nature of the enstrophy
cascade and the slope of the energy
spectrum for 2D flows are not completely settled. Numerical
simulations (e.g., Brachet et al 1988) have suggested that in decaying
(i.e., non-forced) flows, 
distinct evolutionary phases arise. Before small enough scales are
excited, the flow develops ``vorticity shocks'', with a corresponding
$k^{-4}$ spectrum analogous to that of a field of velocity
discontinuities (\S \ref{compressible}). At later stages, once the
dissipative scales have been excited, there appears to be a transition
to a $k^{-3}$ spectrum. Moreover, in 2D incompressible turbulence,
there is a strong tendency towards the formation of long-lived, large-scale
``coherent'' vortices which survive much longer after the small-scale
turbulence has been dissipated by viscosity. The merging of such
vortices can alter the slope of the energy spectrum (see Lesieur 1990,
sect.\ IX.3.2 and references therein). In any case, for {\it forced} 2D
turbulence, the $k^{-3}$ spectrum is the expected equilibrium spectrum, and in
what follows, we will refer to the Kolmogorov and Kraichnan 
spectra as the {\it statistical equilibrium spectra} for 3D and 2D
respectively.

\subsection{Compressible turbulence}\label{compressible}

The compressible case is clearly much more complex than the
incompressible one. The simplest case of a polytropic gas in which the
pressure and the density are related by $P \sim \rho^\gamma$, where
$\gamma$ is the {\it polytropic exponent}, involves already two
coupled, partial differential equations (PDEs) (continuity and
momentum). Full thermodynamics and magnetic fields add one further PDE
each. Self-gravity adds the Poisson equation (see the Virial Theorem
chapter in this book). Thus, self-gravitating, MHD, fully
thermodynamic compressible turbulence may have little to do with the
incompressible case. In the present  section we discuss the
similarities as well as the differences.

\subsubsection{Density fluctuations in isothermal flow.}

A dimensional analysis similar to the one used in the definition of
the Reynolds number in the incompressible case can be used now to
estimate the order of magnitude of the density fluctuations in
isothermal flow, for which $\nabla P \propto c^2 \nabla \rho$, with
$c$ the isothermal sound speed. If the magnitudes of
the pressure and the advective terms are comparable, we find $|\nabla
P| \sim \rho |\u \cdot \nabla \u| \Rightarrow c^2 |\nabla \rho| \sim
\rho U^2/L$, and thus
\begin{equation}\label{rho fluctuations}
{\delta \rho \over \rho} \sim M_a^2,
\end{equation}
where $M_a \equiv U/c$ is the Mach number corresponding to the
characteristic turbulent velocity $U$. 
Note that we have approximated $\nabla \rho$ by
$\delta \rho /L$ and $|\nabla \u|$ by $U/L$, since the mean velocity
of the flow is assumed to be zero. Note also the resemblance of relation
(\ref{rho fluctuations}) with the density jump associated with an
isothermal shock (see, e.g., Shu 1992).

\subsubsection{Compressible and rotational modes.}\label{comp-rot}

In the compressible case, the velocity field is no longer subject to
the incompressibility condition, eq.\ (\ref{incompFS}). However, it is
convenient to still decompose the field in a rotational
(incompressible or solenoidal) part $\u_\k^r$ and a compressible (or
potential) one $\u_k^c$, such that
\begin{equation}\label{comp-roteq}
\k \cdot \u_\k^r =0 \ \ \ \ \ \k \times \u_k^c=0.
\end{equation}

These components of the velocity field can have quite distinct
dynamical properties, and the energy transfer between the two is an
important issue affecting the dynamics of compressible
flows. Interestingly, it appears from numerical simulations of both
moderately (Kida \& Orszag 1990a,b) and highly (\VS\ et al 1996)
compressible turbulence that the transfer is predominantly from
rotational to compressible modes, so that rotational energy tends to
decay if no sources of vorticity $\omega \equiv \nabla \times \u$ are
present. This is particularly interesting since turbulence is normally
thought of as a rotational phenomenon.

\subsubsection{Energy spectrum of compressible turbulence.}

In the highly compressible case, the presence of shocks must also be
considered in the determination of the energy spectrum. A detailed
derivation of the energy spectrum for a field of shocks
can be found, for example, in Saffman
(1971). Here we just give a brief sketch in one dimension. Consider a
shock at the origin 
so that the velocity field can be described by a step function $h(x)$:
$v(x) \propto h(x)$. Then the velocity gradient is proportional to a
Dirac delta function, $dv/dx \propto \delta(x)$, and its Fourier
transform is then a constant, $\F_k(dv/dx) =$ cst., where $\F$ is the
Fourier transform operator. Using the well known result for the Fourier
transform of a derivative, the Fourier transform of the velocity
field is then $v(k)\propto k^{-1}$, and the 1D energy spectrum is finally
$E(k) = v^2(k) \propto k^{-2}$. Thus, the signature spectrum of a
field of 1D shocks has a $-2$ slope. This result is probably related
to observational properties of molecular clouds, as will be seen in \S
\ref{cloud structure}. Note that a $k^{-2}$ spectrum is also
characteristic of Burgers' flows, which are solutions of the one
dimensional {\it Burgers' equation}
\begin{equation}\label{burgers}
{\partial u \over \partial t} + u {\partial u \over \partial x} = \nu
{\partial^2 u \over \partial x^2},
\end{equation}
which also develop ensembles of shocks.
Thus, a $k^{-2}$ spectrum is also known as a Burgers spectrum (see,
e.g., Saffman 1968, sect.\ 6).

Contrary to the incompressible case, for which the Kolmogorov spectrum
is thought to be universal, in the compressible case it appears that
the spectral slope may depend on the degree of compressibility of the
flow. In the weakly compressible regime, 3D numerical simulations indicate
that both components of the velocity develop a statistical equilibrium
(Kolmogorov) spectrum (Porter et al.\ 1994, 1995). At intermediate
compressibilities, 2D simulations indicate that the
compressible component develops a $k^{-2}$ shock spectrum, while the
rotational part maintains a Kraichnan equilibrium spectrum (Passot et al.\ 1988;
\VS\ et al.\ 1995a). At very high compressibility, both components
approach a $k^{-2}$ spectrum (Passot et al.\ 1995). The latter case is
likely to be realized in the ISM, since the density contrast between
the low-density intercloud medium and the densest cloud cores exceeds
seven orders of magnitude.

\subsubsection{Forcing scales and cascades in the ISM.}

A fundamental hypothesis in the K41 theory is that energy is injected
at large scales. Since dissipation occurs predominantly at small
scales, a cascade from large to small scales naturally
arises. However, in the ISM, two important points are that {\it a)} energy
injection occurs at small-to-intermediate scales (Scalo 1987) through
supernova explosions, expanding HII regions, stellar winds and bipolar
outflows, and {\it b)} 
in the compressible case energy can flow between compressible and
rotational modes at each scale (Passot et al.\ 1988), and to internal
energy modes as well. Furthermore, in
the presence of a magnetic field, energy transfer can 
occur to and from magnetic modes. Thus, the concept of a cascade
may not be strictly applicable to the ISM. Nonlinear transfer between
different scales must still occur, but the transfer rate is not
constrained to occur among solenoidal velocity modes exclusively.

\subsubsection{Survival of interstellar turbulence.}

Turbulence is a dissipative phenomenon, and a fundamental question is
whether and how interstellar turbulence can be maintained. This problem is
particularly important in molecular clouds, since in general CO
observations exhibit highly supersonic linewidths for structures
larger than $\sim 0.1$ pc. Since shocks dissipate
much faster than shearing motions (characteristic of incompressible
turbulence), compressible turbulence in molecular clouds 
should decay in roughly the time it takes a strong shock to propagate
across the largest scales (Goldreich \& Kwan 1974). Two principal
solutions have been proposed. First, as mentioned in the previous
paragraph, there are extensive stellar energy sources available for
replenishing the energy. Fleck (1980, 1981) additionally proposed that
shearing due to the Galactic differential rotation could provide the
necessary energy injection. However, this possibility was not
considered to be very
promising (e.g., Shu et al.\ 1987), since it is well known that
{\it Rayleigh's criterion} predicts that a  
rotating fluid is stable if the specific angular momentum decreases
outwards (see, e.g., Shu 1992), so that no turbulence is generated
from the shear. The numerical results of \VS\ et al.\ (1995) support
this result. Nevertheless, it appears that energy injection from
supernovae alone may be enough to maintain the turbulence at large
scales in the ISM (see, e.g., the discussion by Minter \& Spangler 1997).

The second consideration is that, at the scales of  molecular clouds'
interiors, the presence of magnetic fields allows velocities up to the
magnetosonic speed to be present without producing shocks. Since in
general the magnetic field strengths are such that the Alfv\'en speed $\va$
is quite larger than the sound speed $c_s$ (see, e.g.,
Shu et al.\ 1987; Heiles et al.\ 1993), then the magnetosonic
speed $v_m^2 = \va^2 +c_s^2$ is close to the Alfv\'en speed, and the
latter is usually considered as the maximum speed allowed for internal
motions within clouds. In many cases when magnetic field strength
measurememnts are available, the inferred Alfv\'en speeds are indeed
comparable to the velocity dispersions implied by the CO linewidths,
suggesting equipartition between magnetic and kinetic energy (Myers \&
Goodman 1988; Heiles et al 1993), and that indeed magnetic fields
prevent excessive dissipation through shocks.

Note, however, that it is frequently stated (e.g., Shu et al.\ 1987;
Mouscho\-vias 1987) that the regime inside magnetized clouds cannot be
fully turbulent, since in this case case tangled fields should be
observed, while observations tend to indicate smooth field
topologies. This is probably a misconception. If clouds are formed by
turbulent compressions, then the field component perpendicular to the
direction of compression will be amplified due to flux freezing, so
that the field will tend to appear elongated in the same direction as
the cloud (Passot et al.\ 1995). Moreover, turbulent magnetic fields
with normal spectra (say $k^{-2}$, in equipartition with the velocity
at all scales) have {\it more energy at larger scales}, and thus
should appear smooth (i.e., dominated by the larger scales)
anyway. Observations that the magnetic field in clouds generally blends
into the smooth, larger-scale field of the complex in which the
cloud is contained (Goodman et al.\ 1990; see also Heiles et al 1993
sect.\ V) 
are consistent with this interpretation. An alternative
possibility is that the observed magnetic field orientations (using
polarization 
measurements) do not correspond to the field within the clouds, but to
an average along the entire line of
sight to the cloud (e.g., Heiles et al. 1993, sect. I.C.1). In
this case, no explanation in terms of large-scale power is necessary.

\medskip
In summary, the above discussion on compressible turbulence suggests
that it is significantly different from the incompressible case. In
fact, concepts derived form the latter may sometimes not only be
inadequate for the compressible case, but even misleading. In
particular, the concepts of a Kolmogorov spectrum and of an energy
cascade from large to small scales are probably exclusive to
incompressible turbulence, even though they are often considered as
the signatures of the presence of turbulence. For these reasons, it
is probably appropriate to adopt a definition of turbulence based on the
existence of a very large range of excited scales, 
with the ability to nonlinearly exchange excitation among them,
as mentioned in the introduction
(e.g., Lesieur 1990). We now turn to specific applications of
turbulence to molecular clouds.

\section{Applications}\label{applications}

Turbulence is inherently a multi-scale phenomenon, and is thus related
to both large-scale processes such as cloud formation and
morphology, and small-scale processes such as cloud statistical
properties and support against gravitational collapse. We discuss
these below. 

\subsection{Cloud formation}\label{cloud formation}

Cloud formation in the ISM has been a
long-standing problem. A variety of mechanisms have been proposed,
such as coalescence of smaller clouds, various kinds of instabilities
(gravitational, magnetic, thermal, etc.), and turbulence. It appears
that turbulence may be responsible for the formation of the
intermediate and small scale structures (molecular clouds and their
clumps) while the gravitational and magnetic
instabilities may be responsible for the formation of the large-scale
structures (cloud complexes and superclouds) (\VS\ et al.\
1995a,b). In this section we focus on the effects of turbulence. 
Detailed reviews of other processes can be found,
e.g., in Elmegreen (1992, Ch.\ 6) and Balbus (1995) (thermal and gravitational
instabilities), Zweibel (1995 -- magnetic effects) and Elmegreen (1993a
-- coalescence models).

By definition, compressible turbulence implies the formation of
density fluctuations.
Gas compression (shocks, or in general, convergent flows) can  produce
either bound or transient fluctuations, and in particular an important
question is whether structures formed by either turbulent compressions
or passages of single shock waves can become gravitationally unstable
and collpase. In this respect, \"Ogelman \& Maran (1976) and  
Elmegreen \& Lada (1977) suggested that star formation could
be self-propagating, i.e., that shocks induced by supernova explosions
and expanding HII regions could trigger the formation of new
self-gravitating, star-forming clouds. Elmegreen \& Lada (1977) and 
Elmegreen \& Elmegreen (1978)
discussed the gravitational instability of the compressed gas
behind an isothermal shock as a physical mechanism permitting such
self-propagation. Hunter \& Fleck (1982) elaborated on this by noting
that the Jeans mass can be strongly reduced upon turbulent
compression, considering non-isothermal cases. Vishniac (1983) refined
Elmegreen \& Elmegreen's (1978) gravitational instability analysis,
and also discovered a new instability, now known as the Vishniac
instability, through which a shock front with thermal pressure on one
side and bulk (or ``ram'') pressure on the other, tends to
fragment. Vishniac (1994) further considered a {\it nonlinear}
hydrodynamic instability in slabs bounded by shocks on both sides,
showing that this instability can overwhelm the gravitational instability
in the slab. This instability had already been observed in numerical
simulations of shocks between colliding gas streams (Hunter et al.\
1986; Stevens et al.\ 1992) but not properly identified. Elmegreen
(1993) noted that the physical conditions of the medium before
compression may determine whether compressed layers collapse or not,
finding that collapse should occur in molecular clouds, but a
``rebound'' should occur in diffuse clouds.

As pointed out by Hunter et al.\ (1986), when both heating and cooling
are present, the isothermal
approximation, often used to describe radiative flows, is not
appropriate. In cases where the heating and cooling rates are power-law
functions of the density and temperature, the gas can be effectively
described as a
polytropic fluid $P \propto \rho^\gamma$, where $\gamma$ is a
net {\it polytropic exponent} (e.g., Elmegreen 1992; \VS\ et al.\
1996) and is in general different from 1 (Myers 1978). In this case, 
it is well known (Chandrasekhar 1961) that gravitational stability of
a thermally supported spherical cloud requires $\gamma > 4/3$. This
critical value of $\gamma$ becomes 1 in 2D and 0 in 1D. McKee et al.\
(1993) and \VS\ et al.\ (1996) have used related arguments to consider
the stability of fluid parcels compressed in $n$ dimensions by shocks
or turbulence, respectively. The result is that $\gamma$ has to
satisfy $\gamma < 2(1-1/n)$. Shocks are likely to have $n=1$, but
generic turbulent compressions can have any dimensionality $n\leq 3$.

In summary, colliding gas streams
(turbulent or otherwise) are likely to be 
 an efficient mechanism for producing both
self-gravitating and transient clouds. Note that, although
self-gravitating clouds are most notoroius because they are the sites
of star formation, transient clouds are also observed (e.g.,
Carr 1987; Falgarone \& P\'erault 1988; 
Magnani et al.\ 1993) and may even outnumber
self-gravitating ones, albeit containing less mass (\VS\ et al.\ 1997).

\subsection{Cloud structure} \label{cloud structure}

A number of cloud properties are likely to be related to, or even
originated by, turbulence. In this section we discuss Larson's
relations, the clouds' fractal structure and their
velocity distributions.

\subsubsection{Larson's relations}\label{larson's relations}

Larson's (1981) relations have been amply discussed in the Chapter on the
Virial Theorem in this book (herafter VTC). 
Here, we just discuss the various types
of explanations related to turbulence in molecular clouds that have
been proposed for these relations.

For convenience, let us write down Larson's relations again, in their
most commonly accepted form nowadays. They read
\begin{equation}\label{larsoneq}
\Delta v \sim R^{1/2} \ \ \ \ \ \rho \sim R^{-1},
\end{equation}
where $\Delta v$ is the observed velocity dispersion in the cloud
(measured by the linewidth of the observed molecular transitions, or,
for densely sampled maps, by the distribution of the line centroids),
$\rho$ is the cloud's average density, and $R$ is the characteristic
cloud size. The linewidths become  nearly thermal at scales of $\sim
0.03$ pc. Note that, as mentioned in VTC,
these are definitions that originate naturally from poorly
resolved observations, since the cloud often fits within the beam
size, and its edges are blurred, appearing smooth and roughly
circular. However, for high resolution maps, the edges of the best
resolved structures are extremely amorphous, with one projected
dimension in the sky probably quite disparate from the other, making
it difficult to estimate a characteristic ``size''. Nevertheless, we
assume that these quantities are meaningful in order to proceed.

As mentioned in VTC, Larson (1981) originally found slightly
different exponents for the scaling relations. In particular, for the
dispersion-size relation he found an exponent of $0.38$. This exponent
is close to $1/3$, as expected 
for incompressible turbulence, according to the K41
theory (\S \ref{theory}), leading Larson to suggest that the scaling
relation is a manifestation of Kolmogorov turbulence. Subsequent
determinations have pointed towards the exponents
given in relations (\ref{larsoneq}), although  strong
deviations have been found for massive and dense cores (e.g., Caselli
\& Myers 1995; Plume et al 1997), as well as
strongly perturbed regions (e.g., Loren 1989). 

According to the discussion 
in \S \ref{compressible}, theoretically there is no reason
to expect that compressible turbulence should exhibit the same
spectrum (and thus the same scaling of velocity dispersion with size) as
incompressible turbulence, except in the weakly compressible case,
which is {\it not} verified in molecular
clouds. A number of
explanations have been proposed for the 1/2 exponent. One of the
interpretations most
directly related to turbulence is simply that the spectrum expected
for highly compressible turbulence, $E(k) \sim k^{-2}$, appears to
lead naturally to a
$\Delta v \sim R^{1/2}$ scaling relation for the velocity
dispersion. Indeed, {\it assuming that the observed linewidths measure
  the mean square turbulent velocity $u_l$}, we obtain (Passot et al
1988; Padoan 1995; \VS\ \& Gazol 1995; 
Fleck 1996; Gammie \& Ostriker 1996; \VS\ et al.\ 1997)
\begin{equation}
u_l^2 = 2 \int_{2 \pi/l}^\infty E(k) dk \propto \int_{2 \pi/l}^\infty
k^{-2} dk \propto l.
\end{equation}
There are, however, a couple of caveats with this
interpretation. First, the identification of $\Delta
v$ with $u_l$ is not immediate. While the former is a direct measure
of the velocity dispersion, the latter is actually the root mean
square specific kinetic energy in wavenumbers larger than $2 \pi/l$,
and not necessarily a physically measurable quantity. Second, $\Delta v$ is
the velocity dispersion {\it within the beam}, while $u_l$ is an
average over an idealized ensemble (and in practice, over all space;
this is the so-called {\it ergodic hypothesis}). Thus, in order to
decide whether the identification is correct, one should determine
whether the scaling relation is observationally verified also in
diffuse regions of the ISM, i.e., not only at the locations of
density maxima. The
data of Falgarone et al.\ (1992), which include positions in the sky
away from brightness maxima and still exhibit a similar relation
(slope 0.4), seem to support this possibility. 
The simulations of \VS\ et al.\ (1997) seem to also support this
scenario, since 
the dispersion-size relation is verified even when the cloud sample
includes low-density, small clouds.

On a different turbulence viewpoint, Henriksen \& Turner (1984) have
proposed that an inverse cascade {\it of angular momentum} can also
explain both of Larson's relations. Making the two fundamental
assumptions that {\it a)} angular momentum cascades from small to
large scales, so that the torque density is constant through the
inertial range, and that {\it b)} the free-fall time equals the
characteristic time (implying that the characteristic time is the same
for all scales), they are able to derive Larson's
relations. Clearly, the caveat in this approach is that the two
assumptions need  verification. Additionally, as pointed out in \S
\ref{compressible}, compressible turbulence may globally be much less a
rotational regime than incompressible turbulence (\VS\ et al.\ 1996), 
although on the
other hand, vorticity is likely to be generated at small scales in the
ISM, behind curved shocks (e.g., Passot \& Pouquet 1987; 
Fleck 1991), or within thin
shock-bounded slabs (Vishniac 1994).

Explanations based on the transmission of random clump motion to the
ambient medium via Alfv\'en waves have also been proposed. Just et
al.\ (1994) have derived the spectrum of velocity fluctuations induced
on the ambient medium by considering the differential number density
of structures of a given size, from which they can derive the velocity
dispersion-size scaling relation. Finally, another class of
explanations are based on arguments of {\it 
  self-similarity}, which we discuss in the next section, since their
scope is broader than just Larson's relations.

As can be seen from the discussion above, together with the discussion
in VTC, one problem concerning the dispersion-size relation is
that it may be the outcome of a variety of different physical
mechanisms. In fact, many other possible explanations based on
mechanisms othar than the virial theorem or turbulence 
have not been discussed here nor in VTC.
The density-size relation and the cloud mass spectrum are in the same
situation. As of the present time,
a unique theory of the physical processes in clouds
that accounts for all their observed properties is still
lacking.

\subsubsection{Fractal structure}

Molecular clouds are observed to have highly complex structures, manifested in
their morphology, and mass and velocity distributions. 
Clouds are in actuality part of a turbulent continuum with regions of higher
density recursively nested within regions of lower density ({\it
  hierarchical} density structure) (see, e.g., the review by Scalo
1985) and cannot be considered isolated objects, except perhaps at the
smallest, densest scales which decouple from their surroundings due to
self-gravity. Moreover, since the medium is strongly turbulent, the
clouds are naturally amorphous, and their internal velocity distributions need
to be treated statistically. In fact, all of these properties of
clouds appear to be {\it fractal}, i.e., exhibiting substructure at
all scales down to some limit, which may or may have not been yet resolved
(Goodman et al.\ 1997; Minter \& Spangler 1997).

\medskip
\noindent
i) Self-similar models:

Fractal structures are often self-similar; that is, they exhibit the same
substructure properties regardless of scale. Such structures  exhibit
power-law scalings such as the Larson (1981) relations, since it can
be easily shown that a power-law scale is invariant under scale
transformations $x \rightarrow \lambda x$, where $x$ is the spatial
coordinate and $\lambda$ is a scaling factor. Self-similar models for the
ISM (e.g., Ferrini et al.\ (1983); Newman \& Wasserman 1990;
Fleck 1996; Pfenniger 1996) have postulated a
cascade of kinetic energy {\it density} (as opposed to the cascade of
{\it specific} kinetic energy postulated for incompressible flows),
such that its transfer rate $\epsilon_v$ is independent of scale,
analogously to the K41
theory. Additionally, they have postulated a self-similarity law
for the density, so that at each level $n$ of the hierarchy the
density contrast with the previous level is the same. 
Following Fleck (1996), these two postulates read, respectively,
\begin{equation}\label{density cascade}
\epsilon_v = {\rho v^3 \over l} \sim {\rm cst.}
\end{equation}
and
\begin{equation}\label{density similarity}
{\rho_n \over \rho_{n-1}} = \Bigl({l_n \over l_{n-1}}\Bigr)^{-3
  \alpha},
\end{equation}
where $\alpha$ is a ``compression parameter''. Note that
eq.\ (\ref{density similarity}) becomes Larson's density-size relation
for $\alpha = 1/3$. Fleck, however, fixes $\alpha$ by obtaining from
eq.\ (\ref{density cascade}) the form of the energy spectrum as $E(k)
\sim k^{-5/3 - 2 \alpha}$ and equating it to the $k^{-2}$ spectrum for
shock, also appearing in numerical simulations (Passot et al.\ 1995;
Gammie \& Ostriker 1996), obtaining $\alpha=1/6$. Thus, this model
predicts $\rho \sim l^{-1/2}$, in contrast with Larson's relation
$\rho \sim l^{-1}$. Although Fleck attributes the discrepancy to the
neglect of self-gravity in the model,  an important additional
consideration exists. As discussed in \S \ref{compressible}, it is
likely that in ISM turbulent cascades do not properly exist, since
at each scale energy can flow out of kinetic into internal and
magnetic modes. It is thus of fundamental importance to determine the
mechanisms regulating such an exchange. In any case, if the ISM is a
fractal, a number of the observed properties can be explained, such as
the scaling relations, as explained above, the cloud mass spectrum
(Elmegreen \& Falgarone 1996), and other predictions can be made; for
example, ionizing radiation from massive stars could reach to much
larger distances than traditionally thought (Elmegreen 1997)

Another important caveat of self-similar approaches
is that it may be possible that the ISM and
its structures are actually a {\it multifractal}, i.e., a fractal with
different subdivision properties at each level (e.g., Chappel \& Scalo
1996).

\medskip
\noindent
ii) Hierarchical structure. 

Concerning the fractal density structures, an extensive review has
been presented by Scalo (1985). More recently,
Houlahan \& Scalo (1990,
1992) have developed a numerical algorithm capable of distinguishing
between randomly positioned and hierarchically nested structures, and
emphasized on the limitations and distortions of traditional structure
indicators like the correlation function (for the density). 
The method, named ``structure
tree analysis'' by them, amounts to following the branching of the
clouds into smaller subunits as higher densities (or map intensities)
are considered. The method has allowed them to produce a systematic
characterization of interstellar structures by properties such as the
average filling factor of the dense regions, and the mass efficiency
of fragmentation per level of the hierarchy.

A number of mechanisms have been proposed as originators of the
hierarchical structure in the ISM (again, see Scalo 1985 for a
review). In the framework of  
turbulence, it has been normally postulated that, being a self-similar
phenomenon (as indicated by the power-law scalings of the energy
spectrum and the velocity differences -- see \S\ref{incompressible}),
it should produce self-similar density structures as well, as in the
models of Ferrini et al.\ (1983) and Fleck (1996). A first step
towards a mechanism arising from turbulence was suggested by \VS\
(1994), in which the turbulence is assumed to produce density
fluctuations with a specific (but unknown) probability distribution
function (pdf). He suggested that hierarchical structure should arise
naturally provided the pdf $f$ satisfies {\it a)} $f$ does not depend
on the local average density, but only on the fractional density
increment $\delta = d \rho/\rho$, and {\it b)} $f(\delta^n) <
f^n(\delta)$, where 
$n$ is the number of density increment steps to reach a final
density. That is, if the probability of reaching a given large density
increment $\Delta$ is larger when it is reached by a succession of
small steps $\delta$ than when it is reached through a single
density jumps, then one should expect the highest densities to occur
within regions of previously enhanced density. Of course, in this
approach, the physical mechanism determining the specific form of the pdf
is still lacking.

\medskip
\noindent
iii) Fractal cloud boundaries. 

Yet another apparent manifestation of
turbulence in the shaping of clouds is given by their extremely complicated
morphologies. A well known feature of turbulent flows is that they
produce much more efficient mixing than that occurring in laminar
flows. In the incompressible case, such mixing occurs via the
stretching of fluid parcels which intertwine with other parcels due to
shear. Even though in the highly compressible case we have mentioned
the global subdominance of shearing (solenoidal) motions, in shocks
significant vorticity production is expected (Hayes 1957; Passot \&
Pouquet 1987; Fleck 1991; Vishniac 1994), so we should expect complicated 
boundaries in clouds formed by turbulent shock compressions. 

Scalo (1990) and Falgarone et al.\ (1991) have measured the {\it
fractal dimension} of 
a large sample of clouds by measuring their projected areas and
perimeters in the sky. In a simple geometrical object, such as a
circle or a square, the perimeter $P$ is proportional to the square root of the
area $A$. However, fractal objects (e.g., Mandelbrot 1977) that
exhibit substructure at each resolution have perimeters that increase
faster with the area. Indeed, Falgarone et al.\ find that $P \propto
A^{0.68}$, indicating a significant fractality. The only caveat with
this result is that they mention that
the slope of 0.68 is quite similar to that found
for clouds in the Earth's atmosphere, which are subject to very
different physical processes. Thus, it may be that the fractal
dimension is not enough to distinguish between various kinds of flow
regimes.

\medskip
\noindent
iv) Velocity distributions and intermittency.

Clearly, the velocity field is a major protagonist in a turbulent
description of the ISM and molecular clouds. Yet, in the velocity
dispersion-size relationship of Larson's (1981) relations, only the
linewidth of the transition is generally involved, rather than actual
velocity measurements. In fact, linewidths refer to only one observation
per cloud or clump. Several workers have attempted to
improve on this by measuring the actual radial velocities,
obtained from line centroids in densely sampled
maps. Such measurments allow determining statistical indicators
such as the correlation function and correlation length, providing a
characterization of the actual velocity fields in clouds. Scalo (1984)
emphasized these points and attempted measuring the velocity
correlation and structure functions, although he found only very weak
correlations, suggesting a very small correlation length, $\leq 0.05$
pc. Kleiner \& Dickman (1984, 1985) and Dickman \& Kleiner (1985)
presented the theoretical formalism, and performed a detailed study of
the Taurus molecular complex, finding a characteristic length scale for
the density, but still no correlation length for the velocity with a
resolution of 0.6 pc. More recently, Miesch \& Bally (1994) have been
able to measure correlation lengths, albeit noting that the latter are strongly
influenced by the range of scales sampled by the observations, and
conclude that this is a manifestation of the self-similarity of the
flow. 

An important phenomenon known to occur in incompressible turbulence is
{\it intermittency}, consisting of strong, sporadic, 
unpredictable fluctuations in the velocity (see, e.g., Lesieur
1990; Frisch 1995). 
This phenomenon manifests itself in non-Gaussian distributions
of the velocity {\it differences} and {\it derivatives}.
Searches for this behavior have been conducted by Falgarone \&
Phillips (1990) and Miesch and Scalo (1995) for large samples of
clouds. The former authors used line profiles as indicators, while the
latter used again pdfs of the line centroids.  Interestingly, these
groups have found non-Gaussian {\it velocity} statistics, with
exponential distribution tails. This results contrasts with those
found in numerical simulations of compressible turbulence (Passot et al.\ 1995; 
Lis et al.\ 1996), for which the velocities exhibit nearly Gaussian
distributions, while the derivatives exhibit non-Gaussian profiles
(Miesch \& Scalo 1995; 
Lis et al.\ 1996). Thus, the origin of exponential wings in
interstellar velocity distributions is an open problem. Miesch and
Scalo (1995) discuss a number of possible mechanisms, but without
being able to decide among them.

\subsubsection{Cloud support.}

As mentioned in VTC, turbulence at the small scales {\it
  microturbulence}) can provide support against gravitational
collapse, in addition to that provided by thermal pressure and the
magnetic field. In fact, as mentioned before, the linewidths are
generally supersonic, so thermal support is negligible at all but the
smallest scales. Since magnetic support has been discussed at length
in VTC, we restrict ourselves here to a brief review of
the  turbulent support exclusively. 

Turbulent pressure was first considered as an agent of cloud support
by Chandrasekhar (1951), who considered a total pressure of the form
$P=\rho(c^2+\sigma^2/3)$ in the classical Jeans (1902) gravitational
instability analysis. Here $c$ is the sound speed and $\sigma$ is
the turbulent velocity dispersion. This procedure simply enlarges the
Jeans length by a small factor, but gives no qualitatively new
results. Bonazzola et al.\ (1987) recognized that, since the turbulent
energy spectrum indicates that larger eddies contain larger turbulent
energies, the scale dependence should be introduced in the Jeans
analysis. Indeed, they found that for sufficiently steep spectrum
slopes ($E(k) \propto k^{-n}$, with $n\geq 3)$, the Jeans criterion is
{\it reversed}, so that small scales are unstable and large scales are
stable. This in fact is qualitatively correct, since star formation in
the ISM occurs in the smallest cores. However, their required spectral
slopes are too steep, since 3D compressible turbulence is expected to
have slopes $5/3 \leq n \leq 2$ (see \S \ref{compressible}). Just et
al.\ (1994) and \VS\ \&
Gazol (1995) have noted that the Bonazzola et al.\ analysis 
neglects the clumpiness of the ISM, and introduced a power-law
dependence for the density scales. This analysis essentially recovers
the virial equilibrium condition. Just et al., using a particular
model of excitation of MHD waves from clump motions within a cloud,
obtain a $\Delta v \sim R^{1/2}$ relation, and thus
recover the standard exponents for Larson's relations. However, \VS\
\& Gazol take the turbulent spectral index $\alpha$ as a free parameter, and
emphasize that any combination such that $\alpha + \beta =3$, where
$\beta$ is the exponent in the density-size scaling relation, suffices
to achieve virial equilibrium. Thus, Larson's relations are just one
of an infinite number of possible virial equilibrium configurations.
However, the latter authors also emphasize that their analysis clearly
does not apply to all regions in the ISM, but only to density peaks that
follow the specified scaling law for the density. Lower-density
regions away from the peaks necessarily do not follow this scaling relation.
A similar study has been performed by
Vranjes (1994), although
assuming linear, scale-independent density fluctuations.

Turbulent cloud support was tested numerically by L\'eorat et al.\
(1990), who performed  simulations of forced self-gravitating
compressible turbulence and looked for the conditions necessary for
support to be effective. They found that gravitational collapse may be
prevented if the turbulent forcing was injected at sufficiently small
scales. This contrasts with the results of \VS\ et al.\ (1996), in
which gravitational collapse was {\it induced} by turbulent forcing at
{\it large scales}. Thus, turbulence has a twofold character: {\it small
turbulent scales can help prevent gravitational collapse, while large
scales can promote it}. However, the large scale components of the turbulence
are often neglected when considering this problem, and turbulence is
regarded as just providing support, without the ability of
distorting the  cloud morphology and mass distribution.

\subsection{General interstellar turbulence theory}

In this section we wish to describe very briefly some other existing
theoretical work, highlighting their results, as well as their
limitations and, in some cases, flaws. Unfortunately, we have to make the
disclaimer here that this section is by no means complete, but only
presents a random collection of work which we consider of interest.

McKee \& Zweibel (1992) and Zweibel \& McKee (1995) have studied  the
energy budget of clouds, first by producing an Eulerian form of the
Virial Theorem suitable for analysis of turbulent clouds, and then
providing support for the possibility that clouds should  exhibit
energy equipartition. In particular they consider the possibility of
turbulent pressure confinement. However, their results in this respect
are questionable, since they consider confinement by small scale
turbulent modes only, while assuming that large scale modes will only
have the effect of sending shocks through the cloud, thus neglecting
the possibility that the large scale modes actually have the ability
of distorting or disrupting the cloud. Thus, their analysis is
restricted only to the densest cores which have decoupled from their
surroundings due to self-gravity.

Sridhar \& Goldreich (1994) and Goldreich \& Sridhar (1995) have
attempted to produce a theory of interstellar turbulence based on the
assumption that  it consists of nonlinear interactions among shear
Alfv\'en waves. One of the influencing results of this theory is that
the turbulent ``eddies'' should be elongated in the direction of the
magnetic field at small scales.  Such elongations have in fact apparently
been detected (Desal et al.\ 1994). However, the theory has two major
problems. One, it assumes incompressibility, so it is at most
applicable to the diffuse ISM. The same applies to a recent attempt by
Norman \& Ferrara (1997) to characterize a universal forcing function
for interstellar turbulence considering the various types of energy
sources. Two, it has been claimed by Matthaeus
\& Montgomery (1995) that the theory contains a fundamental error
concerning the nature of the nonlinear interactions.

Padoan (1995) has produced a phenomenological theory that predicts
the protostar formation efficiency and mass distribution as functions
of the parameters of the turbulent ambient medium, such as its average density,
temperature, rms turbulent velocity and postshock cooling
time. However, the theory is phenomenological in that it assumes
specific forms for the gas density pdf and the mass spectrum of the
clumps, guided by the results of numerical simulations,
rather than deriving them.

Finally, it is perhaps appropriate to conclude this section noting
that De Vega et al.\ (1996) have pointed out that self-similar
structure could arise in the ISM by purely gravitational effects,
without the need for turbulence.

\section{Conclusions}\label{conclusions}

In this course we have reviewed the basic theory of incompressible and
compressible turbulence, and then attempted to describe the wide
variety of interstellar and molecular cloud phenomena which are likely
to be related to turbulence, in particular the processes of cloud
formation, cloud structuring and cloud support. It should be emphasized
that most treatments are phenomenological, since a full
theory of interstellar turbulence is lacking. Such a theory should be
able to predict properties like the filling factor of the dense
regions, the density pdf, the density power spectrum and correlation
function, and the density contrasts at each hierarchical level (rather
than assuming them); the velocity spectrum and correlation as a
function of the available energy sources, and the efficiency of star
formation. However, since an equivalent theory is not available even
in the incompressible case,  it is clear that interstellar turbulence
research is still in a highly incipient phase.

\vspace*{0.3cm}
\acknowledgements 

The author is glad to acknowledge J.\ Scalo and T.\ Passot 
for helpful comments and a critical reading of the manuscript.



\begin{references}
\reference Armi, L. \& Flament, P. 1985, J. Geophys. Res., 90, no. C6, 11779
\reference Balbus, S. A. 1995, in Physics of the Interstellar Medium and
 Intergalactic Medium, ed. A. Ferrara, C. F. McKee, C. Heiles and
 P. Shapiro (San Francisco: A.S.P), 328
\reference Batchelor, G. K. 1953, The Theory of Homogeneous Turbulence
 (Cambridge: Univ. Press)
\reference Bonazzola, S., Falgarone, E., Heyvaerts, J., P\'erault, M.,
 Puget, J. L. 1987, A\&A 172, 293
\reference Brachet, M. E., Meneguzzi, M., Politano,H. \& Sulem,
P. L. 1988, J. Fluid Mech. 194, 333
\reference Carr, J. S. 1987, ApJ, 323, 170
\reference Caselli, P. \& Myers, P. C. 1995, \apj, 446, 665
\reference Chandrasekhar, S. 1951, Proc. R. Soc. London, 210, 26
\reference Chandrasekhar, S. 1961, Hydrodynamic and Hydromagnetic
 Stability (Oxford: Clarendon Press)
\reference Chappel, D. \& Scalo, J. 1996, preprint
\reference Cowling, T. G. 1976, Magnetohydrodynamics (Bristol: Adam
 Hilger Ltd.)
\reference Currie, I. G., 1974, Fundamental Mechanics of Fluids (New
 York: McGraw-Hill)
\reference Desal, K. M., Gwinn, C. R., \& Diamond, P. J. 1994, Nature,
 372, 754
\reference De Vega, H. J., S\'anchez, N., \& Combes, F. 1996, Nature,
 383, 56
\reference Dickman, R. L. 1985, in  Protostars and Planets II, ed.\
 D. C. Black \& M. S. Matthews (Tucson: Univ. of Arizona Press), 150
\reference Elmegreen, B. G., 1992, in The Galactic Interstellar
 Medium. SAAS-FEE Advanced Course 21, ed. D. Pfenniger and P. Bartholdi
(Berlin: Springer-Verlag), 157
\reference Elmegreen, B. G. 1993a, in Protostars and Planets III,
 ed. E. H. Levy and J. I. Lunine (Tucson: The University of Arizona
 Press), 97
\reference Elmegreen, B. G. 1993b, ApJ, 419, L29
\reference Elmegreen, B. G. \& Lada, C. J. 1977, 1977, ApJ, 214, 725
\reference Elmegreen, B. G. \& Falgarone, E. 1996, ApJ, 471, 816
\reference Elmegreen, B. G. 1997, ApJ, in press
\reference Falgarone, E. \& Puget, J. L. 1986, A\& A, 162, 235
\reference Falgarone, E. \& P\'erault, M. 1988, A\& A, 205, L1
\reference Falgarone, E. \& Phillips, T. 1990, ApJ, 359, 344
\reference Falgarone, E., Phillips, T. G., \& Walker, C. K. 1991, \apj, 378, 186
\reference Falgarone, E., Puget, J.-L., \& P\'erault, M. 1992, A\& A, 257, 715
\reference Fleck, R. C. 1980, ApJ 242, 1019
\reference Fleck, R. C. 1981, ApJ 246, L151
\reference Fleck, R. C. 1991, Ap\&SS, 182, 81
\reference Fleck, R. C. 1996, \apj, 458, 739
\reference Frisch, U. 1995, Turbulence. The Legacy of A. N. Kolmogorov
  (Cambridge: University Press)
\reference Frisch, U. \& Sulem, P. L. 1984, Phys. Fluids, 27, 1921
\reference Gammie, C. \& Ostriker, E. 1996, ApJ, 466, 814
\reference Goldreich, P. \& Kwan, J. 1974, ApJ, 189, 441
\reference Goodman, A. A., Bastien, P., Myers, P. C. \& Menard,
F. 1990, ApJ, 359, 363
\reference Goodman, A. A., Barranco, J. A., Wilner, D. J. \& Heyer,
 M. H. 1997, ApJ, submitted
\reference Grant, H. L., Stewart, R. W. \& Moilliet, A. 1962,
 J. Fluid Mech. 12, 241
\reference Hayes, W. D. 1957, J. Fluid Mech., 2, 595
\reference Heiles, C., Goodman, A., McKee, C. F. \& Zweibel,
 E. G. 1993, in Protostars and Planets III,
 ed. E. H. Levy and J. I. Lunine (Tucson: The University of Arizona
 Press), 279
\reference Henriksen, R. N. \& Turner, B. E. 1984, ApJ, 287, 200
\reference Herring, J. R., Orszag, S. A., Kraichnan, R. H. \& Fox,
 D. G. 1974, J. Fluid Mech., 66, 417
\reference Houlahan, P. \& Scalo, J. 1990, ApJS, 72, 133
\reference Houlahan, P. \& Scalo, J. 1992, ApJ, 393, 172
\reference Hunter, J. H. Jr., \& Fleck, R. C. 1982, ApJ, 256, 505
\reference Hunter, J. H. Jr., Sandford,  M. T. II,  Whitaker, R. W. \&
 Klein, R. I.1986, ApJ, 305, 309 
\reference Jeans, J. H. 1902, Phil. Trans. R. Soc. London, A 199, 1
\reference Just, A., Jacobi, S., \& Deiss, B. M. 1994, A\& A, 289, 237
\reference Kida S., Orszag S.A., 1990a, J. of Sci. Comp., 5, 85
\reference Kida S., Orszag S.A., 1990b, J. of Sci. Comp., 5, 1
\reference Kolmogorov, A. N. 1941, Dokl. Akad. Nauk., 30, 301
\reference Kraichnan, R. H. 1967, Phys. Fluids 10, 1417
\reference Kraichnan, R. H. \& Montgomery, D. 1980, Rep. Prog. Phys.,
 43, 547
\reference Lesieur, M. 1990, Turbulence in Fluids, $2^{\rm nd}$ ed. (Dordrecht:
 Kluwer)
\reference Lis, D. C., Pety, J., Phillips, T. G. \& Falgarone,
 E. 1996, ApJ 463, 623
\reference Magnani, L., LaRosa, T. N., \& Shore, S. N. 1993, \apj,
402, 226
\reference Malerud, S., Maloy, K. J.,\& Goldburg, W.I. 1995,
 Phys. Fluids 7, 1949
\reference Mandelbrot, B. B. 1977, Fractals (San Francisco: Freeman)
\reference McKee, C. F., Zweibel, E. G., Goodman, A.
 A., \& Heiles, C., in Protostars and Planets III, ed. E. H. Levy \&
 J. I. Lunine (Tucson: Univ. of Arizona Press), 327
\reference Miesch, M. S. \& Bally, J. 1994, \apj, 429, 645
\reference Miesch, M. S. \& Scalo, J. 1995, ApJ, 450, L27
\reference Minter \& Spangler 1977, ApJ submitted
\reference Mouschovias, T. Ch. 1987, in Physical Processes in
 Interstellar Clouds, ed. G. E. Morfill and M. Scholer (Dordrecht: Reidel), 453
\reference Myers, P. C. 1978, ApJ, 225, 380
\reference Myers, P. C., \& Goodman, A. A. 1988, \apj, 326, L27
\reference Newman, W. I., \& Wasserman, I. 1990, \apj, 354, 411
\reference Norman, C. \& Ferrara, A. 1997, preprint
\reference Obukhov, A. M. 1941, C. R. Acad. Sci. USSR, 32, 19
\reference \"Ogelman, H. B. \& Maran, S. P. 1976, ApJ, 209, 124
\reference Ott, E., 1993, Chaos in Dynamical Systems (Cambridge:
 University Press)
\reference Padoan, P. 1995, MNRAS, 277, 377
\reference Passot, T., \& Pouquet, A. 1987, J. Fluid Mech., 181, 441
\reference Passot, T., Pouquet, A., \& Woodward, P. R. 1988, A\&A, 197, 228
\reference Passot T., V\'azquez-Semadeni E., Pouquet A., 1995, ApJ, 455, 702
\reference Pfenniger, D. 1996, preprint
\reference Porter, D. H., Pouquet, A., Woodward, P. R. 1994,
 Phys. Fluids, 6, 2133
\reference Porter, D. H.,
 Pouquet, A., Woodward, P. R. 1995, in Small-Scale Structures in
 Three-Dimensional Hydrodynamic and Magnetohydrodynamic Turbulence, ed.
 M. Meneguzzi, A. Pouquet, P.-L. Sulem (Berlin: Springer), 51
\reference Rose, H. A. \& Sulem, P. L. 1978, J. Physique, 39, 441
\reference Saffman, P. G. 1968, in Topics in Nonlinear Physics,
ed. N. J. Zabusky (New York: Springer-Verlag), 485
\reference Saffman, P. G. 1971, Stu. Appl. Math., 50, 377
\reference Scalo, J. 1984, ApJ, 277, 556
\reference Scalo, J. 1985, in Protostars and Planets II, ed. D. C. Black and
 M. S. Matthews (Tucson: Univ. of Arizona Press), 201
\reference Scalo, J. M. 1987, in Interstellar Processes, ed.
 D. J. Hollenbach and H. A. Thronson (Dordrecht: Reidel), 349
\reference Scalo, J. 1990, in Physical Processes in Fragmentation and Star
 Formation, ed. R. Capuzzo-Dolcetta, C. Chiosi, \& A. di Fazio
 (Dordrecht: Kluwer), 151
\reference Shore, S. N. 1992, An Introduction to Astrophysical
 Hydrodynamics (San Diego: Academic Press)
\reference Shu, F. 1992, Gas Dynamics (Mill Valley: University Science
 Books)
\reference Shu, F., Adams, F. C., Lizano, S. 1987, ARAA, 25, 23
\reference Spangler, S. R. 1991, ApJ, 376, 540
\reference Spitzer, L., Jr., 1978, Physical Processes in the
 Interstellar Medium (New York: Wiley-Interscience)
\reference Stevens, I. R., Blondin, J. M \& Pollock, A. M. T. 1992,
 ApJ 386, 265
\reference Tennekes, H. \&  Lumley, J. L. 1972, A First Course in
 Turbulence (Cambridge: MIT Press)
\reference V\'azquez-Semadeni, E. 1994, ApJ, 423, 681
\reference V\'azquez-Semadeni, E., Passot T., Pouquet A., 1995a, ApJ, 441, 536
\reference V\'azquez-Semadeni, E.,  Passot T., Pouquet A., 1995, in
 Proceedings of the Fifth Mexico-Texas Conference on Astrophysics:
 Gaseous Nebulae and Star Formation ,  Rev. Mex. A.A. Ser. Conf., 3, 61.
\reference V\'azquez-Semadeni, E., Passot T., Pouquet A., 1996, ApJ in
 press (Dec. 20)
\reference \VS, E., \BP, J. \& Rodr\'\i guez, L. F. 1997, ApJ, in
 press (Jan. 1)
\reference Vranjes, J. 1994, Ap\& SS, 213, 139
\reference Zweibel, E. 1995, in Physics of the Interstellar Medium and
 Intergalactic Medium, ed. A. Ferrara, C. F. McKee, C. Heiles and 
 P. Shapiro (San Francisco: A.S.P), 524
\end{references}
\end{document}